# Fusing Structural Phenotypes with Functional Data for Early Prediction of Primary Angle-Closure Glaucoma Progression

Swati Sharma, PhD,[1] Thanadet Chuangsuwanich, PhD,[3] Royston K.Y. Tan, PhD,[1,2] Shimna C. Prasad, MS,[1] Tin A. Tun, MD, PhD,[1,2] Shamira A. Perera, MBBS, FRCOphth,[1,2] Martin L. Buist, PhD,[4] Tin Aung, FRCSEd, PhD,[1,2] Monisha E. Nongpiur, MD, PhD,[1,2] Michaël J. A. Girard, MSc, PhD,[1,2,3,5,6]

1. Singapore Eye Research Institute, Singapore National Eye Centre, Singapore
2. Duke-NUS Graduate Medical School, Singapore
3. Department of Ophthalmology, Emory University School of Medicine, Emory University
4. Department of Biomedical Engineering, National University of Singapore, Singapore, Republic of Singapore
5. Department of Biomedical Engineering, Georgia Institute of Technology/Emory University, Atlanta, GA, USA
6. Emory Empathetic AI for Health Institute, Emory University, Atlanta, GA, USA

**Corresponding Authors**:

Michaël J.A. Girard
Ophthalmic Engineering & Innovation Laboratory
Emory Eye Center, Emory School of Medicine
Emory Clinic Building B, 1365B Clifton Road, NE
Atlanta GA 30322
mgirard@ophthalmic.engineering
Monisha Esther Nongpiur
Singapore Eye Research Institute, The Academia, 20 College Road, Discovery Tower Level 6, 169856, Singapore
monisha.esther.nongpiur@seri.com.sg

**Abbreviations:**

AUC: Area Under the Receiver Operating Characteristic Curve

BMO: Bruch's Membrane Opening

GCC: Ganglion Cell Complex

GCL+IPL: Ganglion Cell-Inner Plexiform Layer Complex

HFA: Humphrey Field Analyzer

I: Inferior

IN: Inferior-Nasal

IOP: Intraocular Pressure

IT: Inferior-Temporal

LC: Lamina Cribrosa

MD: Mean Deviation

ML: Machine Learning

MRW: Minimum Rim Width

N: Nasal

OCT: Optical Coherence Tomography

ONH: Optic Nerve Head

ORL: Other Retinal Layers

PACG: Primary Angle Closure Glaucoma

POAG: Primary Open Angle Glaucoma

RGC: Retinal Ganglion Cells

RNFL: Retinal Nerve Fiber Layer

ROC: Receiver Operating Characteristic Curve

RPE: Retinal Pigment Epithelium

S: Superior

SD-OCT: Spectral-Domain Optical Coherence Tomography

SHAP: Shapley Additive Explanations

SITA: Swedish interactive threshold algorithm

SN: Superior-Nasal

SNEC: Singapore National Eye Centre

ST: Superior-Temporal

T: Temporal

VF: Visual Field

VFI: Visual Field Index

**Abstract**

**Purpose:** To classify eyes as slow or fast glaucoma progressors in patients with primary angle closure glaucoma (PACG) using an integrated approach combining optic nerve head (ONH) structural features and sector-based visual field (VF) functional parameters.

**Design:** Retrospective longitudinal study

**Participants:** PACG patients from glaucoma clinics

**Methods:** PACG patients with ≥5 reliable VF tests over ≥5 years were included. Progression was assessed in Zeiss Forum, with baseline VF within six months of OCT. Fast progression was VFI decline <-2.0% per year; slow progression ≥-2.0% per year. OCT volumes were AI-segmented to extract 31 ONH parameters. The Glaucoma Hemifield Test defined five regions per hemifield, aligned with RNFL distribution. Mean sensitivity per region was combined with structural parameters to train ML classifiers. Multiple models were tested, and SHAP identified key predictors.

**Main outcome measures:** Classification of slow versus fast progressors using combined structural and functional data.

**Results:** We analyzed 451 eyes from 299 patients. Mean VFI progression was -0.92% per year; 369 eyes progressed slowly and 82 rapidly. The Random Forest model combining structural and functional features achieved the best performance (AUC = 0.87±0.02, 2000 Monte Carlo iterations). SHAP identified six key predictors: inferior MRW, inferior and inferior-temporal RNFL thickness, nasal-temporal LC curvature, superior nasal VF sensitivity, and inferior RNFL and GCL+IPL thickness. Models using only structural or functional features performed worse with AUC of 0.82±0.03 and 0.78±0.03, respectively.

**Conclusions:** Combining ONH structural and VF functional parameters significantly improves classification of progression risk in PACG. Inferior ONH features, MRW and RNFL thickness, were the most predictive, highlighting the critical role of ONH morphology in monitoring disease progression.

**Introduction**

Primary angle closure glaucoma (PACG) is a major subtype of glaucoma characterized by a mechanical obstruction of the anterior chamber angle, resulting in impaired aqueous humour outflow and elevated intraocular pressure (IOP),[1] which can progressively lead to raised IOP and subsequent optic nerve damage and irreversible vision loss.[2] The chronic form of PACG may remain asymptomatic until the late stages, when significant VF loss has already occurred, making early detection particularly challenging. Although PACG is less prevalent than primary open angle glaucoma (POAG), epidemiological studies have shown that it causes disproportionately higher rate of blindness. This underscores the importance of timely identification of at-risk individuals and close monitoring of both structural and functional parameters. Early prediction of disease progression using biometric, structural, and clinical indicators could enable proactive management strategies. There is, thus is a pressing need for predictive models to assess progression risk in PACG, guiding clinicians in preserving visual function and optimize patient outcomes.

Standard automated perimetry is the most widely used functional test for monitoring glaucoma progression and is considered the clinical gold standard for evaluating visual function. Numerous research efforts have focused on predicting future VF decline based on existing VF data.[3,4] However, due to the subjective nature and significant variability of VF measurements, reliable predictions usually require a series of 6 to 8 tests. Consequently, it may take approximately 3 to 4 years following the initial diagnosis to determine the true rate of progression. In clinical practice, obtaining such a large number of VFs is often impractical, as treatment decisions frequently need to be made after only a few visits, making it challenging to identify fast progressors accurately and at an early stage. Further complicating the assessment, variability in VF outcomes may result from various factors, such as coexisting cataracts, advanced glaucoma with frequent fixation losses, learning effects during testing, or even patient distraction.

Glaucoma progression can be evaluated by analyzing both structural and functional parameters. In some cases, structural damage can be identified before functional impairment using current diagnostic tools, making it a potential predictor of future visual changes. Optical coherence tomography (OCT) imaging of the ONH provides valuable structural information, including measurements related to the retinal nerve fiber layer (RNFL) and the ganglion cell-inner plexiform layer (GCL+IPL) complex. To address the variability of VF data and enhance the accuracy of glaucoma detection, several studies have integrated these structural data with functional metrics. For example, Garway-Heath et al. demonstrated that combining VF and OCT data allowed for more precise estimation of glaucoma progression rates compared to using VF data alone.[5] Similarly, Dixit et al. showed that incorporating basic clinical data (such as cup-to-disc ratio, corneal thickness, and intraocular pressure) alongside VF data improved the model's predictive performance.[6] Despite achieving good accuracy, these studies faced limitations in explainability, as they were unable to identify the specific factors driving the predicted outcomes.

This study aimed to integrate structural features of the ONH, derived from OCT images using AI-based 3D structural analysis, with functional features from VF data to enable early prediction of glaucoma progression. The specific goals were: 1) to extract structural parameters of the ONH from OCT images and combine them with VF data from five distinct sectors in superior and inferior hemifields; 2) to develop machine learning (ML) models to classify eyes as slow or fast progressors; and 3) to use Shapley Additive Explanations (SHAP) to identify the most influential factors driving progression and understand their relationship with glaucoma progression.

**Methods**

**Patient recruitment and visual field progression**

The retrospective study was conducted by reviewing the medical records of patients diagnosed with PACG who were seen at the glaucoma clinics of the Singapore National Eye Centre (SNEC) between 2005 and 2016. These included patients had previously participated in a research study focused on identifying genetic factors associated with PACG.[7,8] At the time of biological sample collection, written informed consent was obtained from all participants for access to their clinical data. Ethical approval for the study was granted by the SingHealth Centralized Institutional Review Board and was conducted in adherence to the principles outlined in the Declaration of Helsinki. Table 1 summarizes the subjects population.

| Study | Sex (% of female) | Age (mean± SD) | Mild | Moderate | Total |
|---|---|---|---|---|---|
| Cohort | 52 | 67±7.7 | 312 | 139 | 451 |

*Table 1: Subject demographics and number of subjects in each category*

PACG was defined as the presence of angle closure (where ≥180° of the posterior pigmented trabecular meshwork was not visible on non-indentation gonioscopy in the primary gaze position) with glaucomatous optic neuropathy and compatible VF defect. Patients were excluded if they had any other ocular condition in the study eye that could account for the VF loss other than glaucoma.[7,8]

Visual field testing was conducted using the 24-2 Swedish interactive threshold algorithm (SITA) strategy (either SITA Standard or SITA-FAST) with the Humphrey Field Analyzer (Carl Zeiss Meditec, Dublin, California) with appropriate refractive correction. The Zeiss Forum platform was utilized to assess longitudinal VF changes, with the software allowing integration of SITA Standard and SITA Fast tests to provide the overall slope of progression as visual field index (VFI), percentage (%) per year, as well as the superior and inferior slopes of progression. To align VF progression analysis with structural imaging, the baseline VF test in the Forum software was re-selected to fall within six months of the OCT scan. Patients who had at least five or more reliable visual fields and at least five years follow-up were evaluated. Fast progression was defined as a VFI decline of < -2.0% per year, while slow progression was defined as a decline of ≥ -2.0% per year. Disease severity was classified based on the first visual field test used in the Forum software, categorized by mean deviation as mild (≥-6.0dB), moderate (-6.01 to -12.0dB) or severe (<-12.01dB).[8] Participants with severe VF loss were excluded, as glaucoma progression tends to be slower in such cases due to the floor effect in VF assessment,[9] and because structural-functional correlations become less reliable in advanced disease stages. Raw retinal sensitivity values at individual test points were extracted from VF printouts obtained within 6 months of the OCT scan. To ensure spatial alignment, VF maps from left eyes were horizontally flipped to match the configuration of right eyes. For regional VF analysis, five distinct sectors corresponding to specific retinal nerve fiber layer (RNFL) anatomical features were defined within each hemifield along the horizontal meridian. These sectors included the central, paracentral, nasal, arcuate 1, and arcuate 2 regions in both the superior and inferior hemifields. This segmentation was designed to emphasize the asymmetric field loss patterns commonly observed in glaucoma.[8] The VF sensitivity values within the ten sectors were averaged to obtain a single representative VF sensitivity value for that sector.

**Spectral-domain optical coherence tomography imaging**

Spectral-domain optical coherence tomography (SD-OCT) imaging was conducted using the CIRRUS HD-OCT 6000 system (Carl Zeiss Meditec, Dublin, CA), employing its standard 6x6 $mm^2$ scan protocol centered on the optic nerve head. This protocol captures a data cube over a 6-mm square grid by acquiring 200 horizontal scan lines, each containing 200 A-scans. All scans were manually inspected using Zeiss CIRRUS HD-OCT Review Software 11.5.1. Scans of suboptimal quality, identified by segmentation errors, uneven signal intensity, excessively weak signal strength, or significant motion artifacts, were excluded from the analysis.

**ONH morphological parameter extraction**

The Reflectivity software (Abyss Processing Pte Ltd, Singapore) was used to segment the retinal and connective tissue layers of the ONH from each OCT volume (see Fig. 1). These layers included the RNFL, the ganglion cell layer combined with the inner plexiform layer (GCL+IPL), the remaining other retinal layers (ORL), the retinal pigment epithelium (RPE), the choroid, the visible part of the peripapillary sclera (including the scleral flange), and the lamina cribrosa (LC). The software was also used to extract 31 structural parameters, such as Bruch's Membrane Opening (BMO) area, LC curvature and depth, prelamina depth and thickness, scleral angle, and average thickness of the RNFL layer, GCL+IPL layer, and Minimum Rim Width (MRW) across the superior (S), superior-temporal (ST), superior-nasal (SN), nasal (N), temporal (T), inferior-nasal (IN), inferior-temporal (IT), and inferior (I) regions. Detailed definitions of these parameters are provided in the supplemental material.

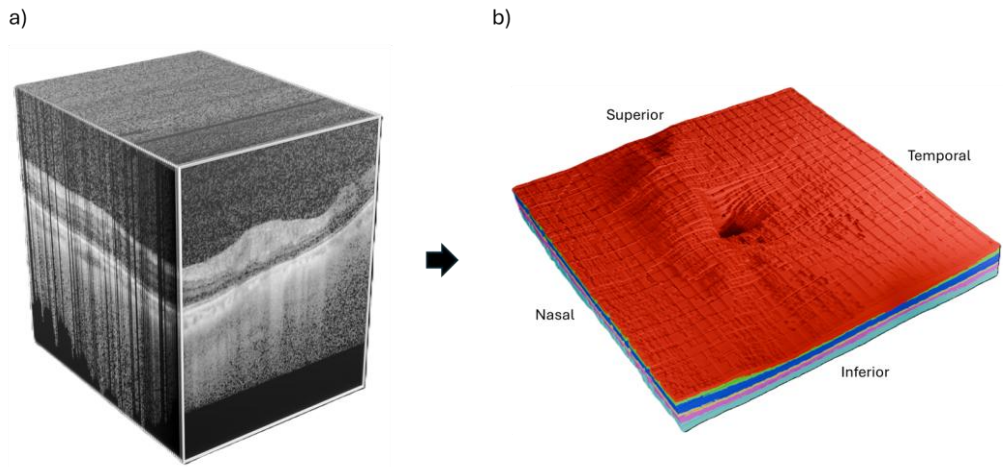

*Figure 1: a) A sample OCT volume of the optic nerve head. b) The ONH tissue layers segmented using Reflectivity and rendered using Blender.*

**Machine learning approach for VF progression prediction**

In addition to the 31 structural parameters derived from OCT images, 10 functional parameters extracted from VF maps, i.e., the mean retinal sensitivity values across the 10 visual field sectors were included as inputs for various ML models (Fig. 2). These models were trained to distinguish between eyes with slow and fast progression. The ML models tested for this classification task included tree-based machine learning models, such as Random Forest, Decision Tree and Extra Trees classifier, Gradient Boosting algorithms, such as XGBoost, CatBoost, and HistGradientBoosting classifier, and neural network based Multi-Layer Perceptron classifier. These algorithms are widely recognized for their effectiveness in healthcare applications.[10–12] Furthermore, to evaluate the individual contributions of structural and functional parameters, we conducted two separate simulations using each parameter set independently and compared their classification performance.

Due to the dataset's imbalance, with the slow progression class being approximately four times larger than the fast progression class, higher class weights were assigned to the fast progression group. The class weights were computed in such a way that the weight for each class was inversely proportional to its frequency. This adjustment mitigated the risk of model bias toward the slow progression class during training.

To assess model reliability, Monte Carlo cross-validation was employed.[13] The dataset was randomly partitioned into 70% training, 15% validation, and 15% testing data over 2000 iterations. In each iteration, the ML models were trained afresh on the training data, and their performance was assessed on the test set. The area under the receiver operating characteristic (ROC) curve (AUC) was calculated in each iteration to evaluate model accuracy. Additionally, accuracy, sensitivity, and specificity values were reported using the confusion matrix.

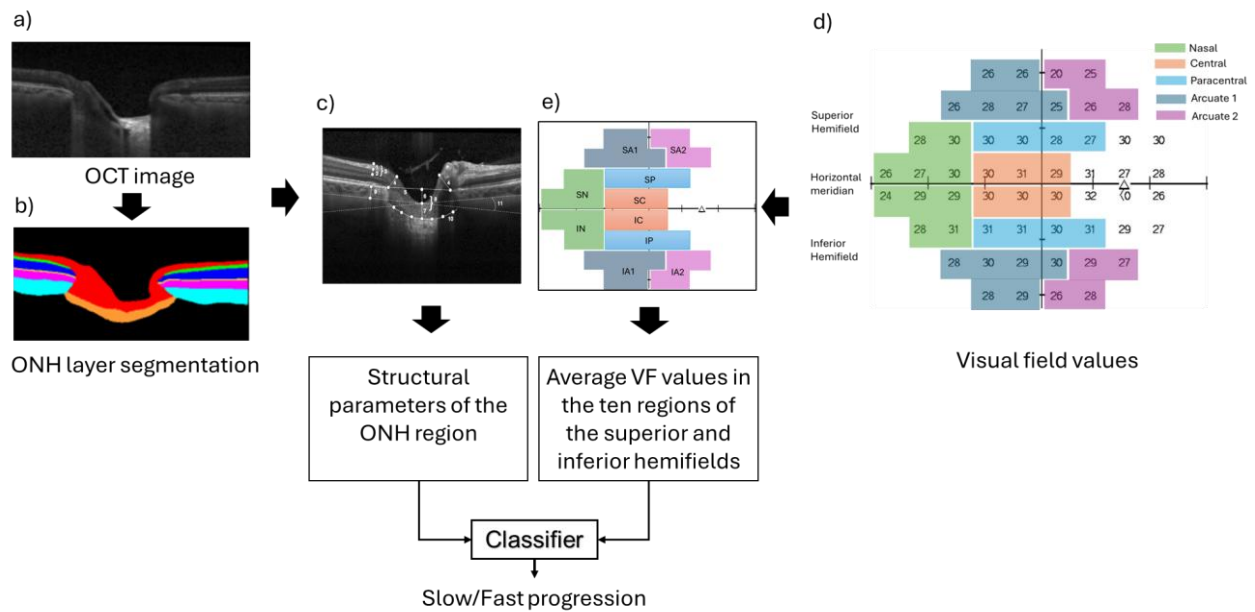

*Figure 2: The proposed framework for slow and fast progression detection using the structural parameters of the ONH from OCT images and the functional parameters from VF maps. Five distinct sectors corresponding to specific RNFL anatomical features were defined within each hemifield of VF map along the horizontal meridian. These sectors included the central (orange), paracentral (blue), nasal (green), arcuate 1 (grey), and arcuate 2 (purple) regions in the superior and inferior hemifields.*

**Explainability of ML approach**

We employed SHAP to interpret the model's predictions and enhance explainability.[14] SHAP offers a comprehensive framework for understanding ML model outputs by utilizing Shapley values from cooperative game theory. This method equitably distributes the prediction outcome among all features, ensuring that each feature's contribution is accurately assessed. By assigning each feature a SHAP value that reflects its influence on the model's output, the method provides consistent and interpretable explanations.

The "TreeExplainer" method from SHAP was utilized to identify key features in each iteration of the Monte Carlo cross-validation process. These feature importance scores were averaged across 2000 iterations to obtain the mean SHAP values and overall feature importance. This averaging technique enabled us to derive a stable estimate of feature significance throughout the repeated validation process.

**Results**

A total of 451 eyes from 299 patients were evaluated. The mean age at the time of OCT scan was 67±7.7 years, and 52% were females. Of the 451 eyes, 312 had mild VF loss and 139 had moderate VF loss. The mean rate of VFI progression was -0.92% per year; 369 eyes demonstrated slow progression, while 82 exhibited fast progression.

Our ML models, which integrated structural and functional parameters, demonstrated high accuracy in distinguishing between slow and fast progressing eyes. Among all evaluated models, the Random Forest algorithm achieved the best performance on an independent test set, with an AUC of 0.87 ± 0.02 over 2000 Monte Carlo cross-validation iterations. Gradient Boosting models also delivered comparable results. A summary of performance metrics for all ML models is presented in Table 2. The highest AUC recorded was 0.89, wherein the model correctly identified 59 of 68 slow-progressing eyes and 19 of 23 fast-progressing eyes (see Fig. 3a and 3b).

| ML method | AUC (Monte Carlo cross-validation with 2000 iterations) |
| --- | --- |
| Random Forest Classifier | 0.87±0.02 |
| Decision Tree Classifier | 0.77±0.04 |
| Extra Trees Classifier | 0.84±0.03 |
| XGBoost Classifier | 0.85±0.02 |
| CatBoost Classifier | 0.86±0.03 |
| HistGradientBoosting Classifier | 0.86±0.02 |
| MLP classifier | 0.80±0.02 |

*Table 2: Comparative performance of machine learning models for predicting slow and fast progression*

The top six features contributing to the classification as identified by SHAP analysis were: MRW in the inferior region, average RNFL thickness in the inferior region, nasal-temporal LC curvature, average VF sensitivity in the superior nasal sector, RNFL thickness in the inferior-temporal region, and GCL+IPL thickness in the inferior region. The top fifteen most influential parameters based on SHAP values are shown in Fig. 3c.

Models combining both structural and functional parameters outperformed those using either alone, with AUCs of 0.82±0.03 (structural alone) and 0.78±0.03 (functional alone), as shown in Table 3. SHAP analysis on structural-only and functional-only data revealed key features consistent with those identified in the combined dataset. Fig. 4a and Fig. 4b display the most important parameters from the structural-only and functional-only simulations, respectively.

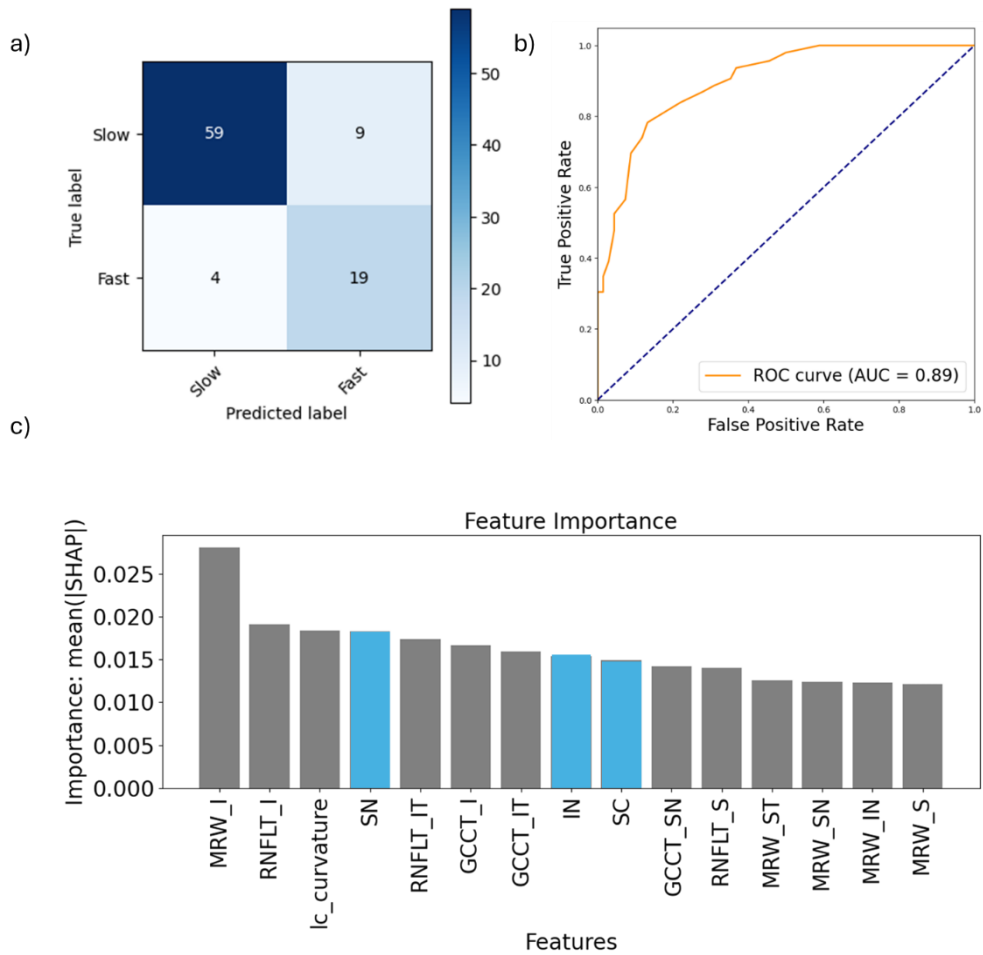

Figure 3: a, b) Confusion matrix showing the numbers of correctly classified eyes in each class and the ROC curve for the best classification result. c) Structural (grey bars) and functional parameters (blue bars) arranged in terms of their importance for slow and fast progression classification (See supplementary material for the definition of the parameters).

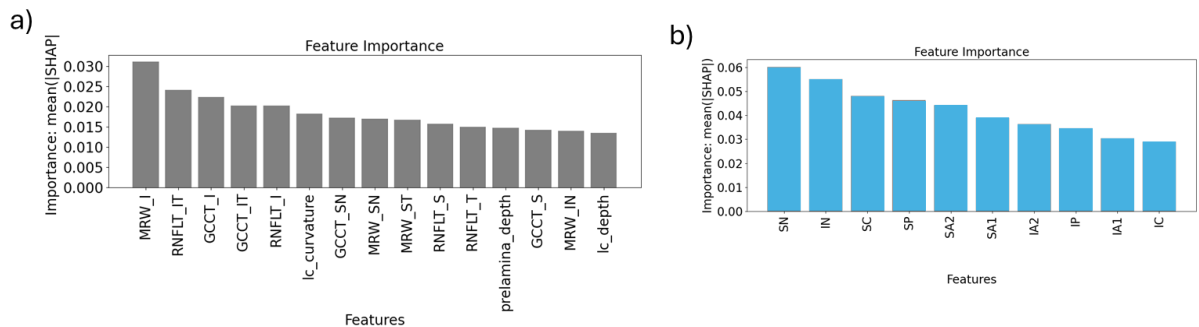

Figure 4: a) Structural features in terms of their importance for slow and fast progression classification. b) Functional features in terms of their importance.

| ML method | AUC with structural and functional parameters combined | AUC with structural parameters only | AUC with functional parameters only |
|---|---|---|---|
| Random Forest Classifier | 0.87±0.02 | 0.82±0.03 | 0.78±0.03 |

*Table 3: Accuracy of predicting slow and fast progression with only structural parameters, functional parameters, and both combined*

Dependency plots revealed strong correlations between input features and their SHAP values. For example, as shown in Fig. 5a, inferior RNFL thickness exhibited a negative correlation, with decreasing thickness linked to increasing SHAP values, indicating a higher risk of fast progression. All features except LC curvature followed this trend. In contrast, the nasal-temporal LC curvature showed a positive correlation, where increased curvature corresponded with higher SHAP values and a greater probability of fast progression.

**Discussion**

In this study, we integrated structural parameters of the ONH derived from OCT images with functional parameters from VF assessments to evaluate their combined effectiveness in classifying eyes as slow or fast progressors. The analysis was carried out in two stages. First, multiple ML models were applied to assess classification performance. Second, SHAP analysis was used to identify the most influential parameters contributing to the classification. The ML models demonstrated high accuracy, and the SHAP analysis effectively highlighted the key predictive features along with their correlations with slow and fast progression in PACG.

Several studies have investigated the mechanisms underlying glaucoma progression primarily focusing on POAG due to its higher prevalence in studied populations.[4,15,16] Although PACG is nearly three times less common than POAG, the number of individuals who go blind from PACG is comparable.[17] This may be attributed to a limited understanding of PACG progression and a lack of optimized treatment strategies. PACG is particularly prevalent in Asian populations, accounting for approximately 86% of the global angle-closure glaucoma burden.[18] Despite this, literature on PACG progression remains limited, with most studies emphasizing VF data.[19–22] Recent research has begun to highlight the significance of RNFL and GCC thickness measurements from OCT in tracking PACG progression.[23,24] In this study, we designed a novel ML- and SHAP-based tool to predict PACG progression using only baseline data from one of the largest PACG cohorts to date. To our knowledge, this is the first approach to integrate functional VF data with structural ONH features from OCT for PACG progression prediction and to identify key features of disease progression. This framework also holds potential for application in POAG progression analysis.

Our proposed model demonstrated high accuracy in distinguishing slow-progressing eyes from rapidly progressing ones with an AUC of 0.87 ± 0.02. Previous research by Shuldiner et al., Garway-Heath et al., and Dixit et al. have highlighted the predictive value of VF maps and clinical parameters, such as cup-to-disc ratio, corneal thickness, and IOP in assessing glaucoma progression.[3,5,6] However, these studies primarily focused on POAG, and it remains unclear whether these findings can be extrapolated to PACG, given the differences in clinical presentation, treatment approaches, and risk factors. Our study, focused exclusively on PACG, provides compelling evidence that integrating structural features from OCT images with functional insights from VF maps presents a promising new approach to tracking glaucoma progression.

Our SHAP-based analysis identified key ONH structural parameters, such as MRW, average RNFL and GCL+IPL thickness in the inferior region, as well as RNFL thickness in the inferior-temporal region, as critical predictors of disease progression (Fig. 3c and Fig. 4a). The SHAP dependency plots revealed a negative correlation between SHAP values and these structural parameters, indicating that lower values of these parameters are associated with faster disease progression (Fig. 5a–5d). These findings also build on our earlier cross-sectional observations in PACG, which showed that superior hemifield defects were more pronounced than inferior ones[8]. In the current study, we further demonstrate that these superior defects are more closely associated with disease progression. This is anatomically consistent, as structural damage in the inferior ONH corresponds to superior hemifield VF loss.

Our study demonstrated that the superior hemifield sectors play a more significant role in predicting disease progression compared to the inferior ones (see Fig. 3b and Fig. 4b). This observation aligns with several prior studies. For example, Verma et al. reported that superior sectors showed earlier progression in early-stage PACG,[19] while multiple cross-sectional studies have also documented greater total deviation values in the superior hemifield than in the inferior hemifield.[19,25] Yousefi et al. similarly observed more prominent VF damage in the superior hemifield.[26] These findings suggest that nasal and peripheral VF regions particularly in the superior hemifield, are more susceptible to early glaucomatous damage, which may later evolve into diffuse loss as the disease progresses.

Anatomically, this pattern is consistent with the known structural vulnerability of the inferior optic nerve head, which has thinner RNFL bundles and is more prone to pressure-related damage.[5] Given that PACG is a pressure-dependent form of glaucoma, chronic or sub-acute IOP elevation often results in focal or diffuse RNFL loss, particularly in the inferior retina, manifesting as superior hemifield defects on VF testing.

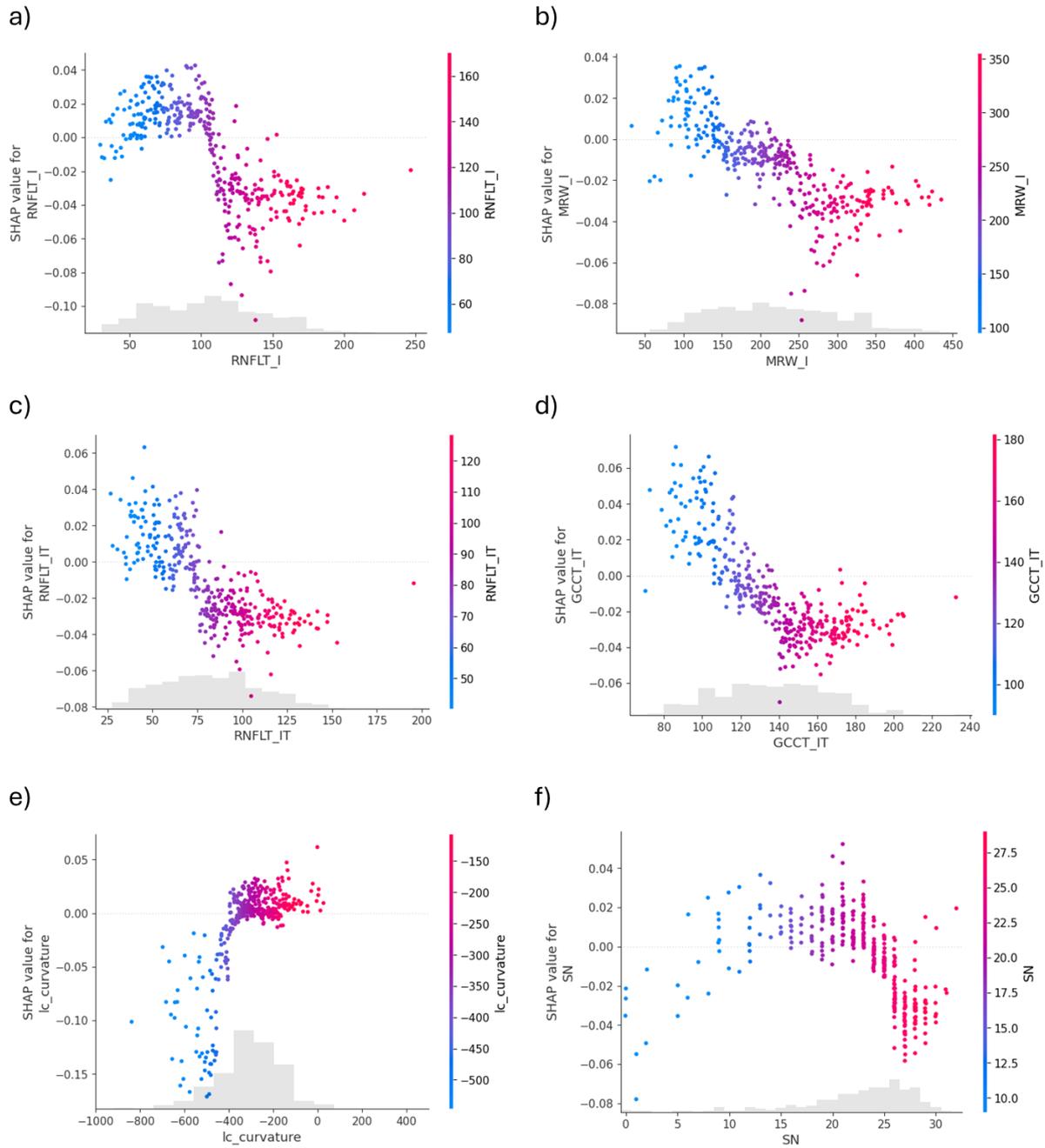

*Figure 5: Dependency plots illustrate the relationship between the most influential structural and functional parameters and their corresponding SHAP values. a-d) A negative correlation between SHAP values and RNFL thickness, GCL+IPL thickness, and MRW, indicating that lower values of these parameters are associated with faster disease progression. e) A positive correlation between LC curvature and SHAP values, suggesting that increased LC curvature is linked to faster progression. f) A negative correlation between average VF sensitivity and SHAP values, indicating that reduced VF sensitivity in this sector is associated with faster progression.*

Our SHAP-based analysis independently confirmed this structure-function relationship, highlighting that several superior VF sectors, specifically the superior nasal, central, and paracentral sectors were among the most critical in predicting progression. Notably, the inferior nasal sector also emerged as an important contributor, although superior hemifield regions were more prominently represented overall.

Our findings suggest that MRW may serve as key structural marker for progression in PACG. Sihota et al. observed significant sectoral ONH changes in PACG, likely resulting from localized ischemia caused by subacute IOP spikes.[27] Chauhan et al. first introduced MRW as a robust structural parameter aligned with the true anatomical path of axonal exit.[28] Subsequent studies have shown that MRW is significantly reduced in glaucomatous eyes, correlating strongly with RNFL loss and VF deterioration.[29,30] The findings from these studies are in agreement with our observations.

We observed that RNFL and GCL+IPL thinning may be a critical structural biomarker for PACG progression. Many studies reported that progressive RNFL thinning mirrors VF loss in PACG. Aung et al. reported that acute PACG Initially presents with transient RNFL thickening due to optic disc edema, followed by significant thinning that reflects irreversible axonal damage.[31] Other studies have reported a strong correlation between the VF progression and thinning of the GCC and RNFL.[24,32] Our study reinforces these findings, emphasizing the role of RNFL and GCL+IPL thinning in PACG progression.

Our study also identified LC curvature as a structural parameter associated with disease progression in PACG. A positive correlation was observed between nasal temporal LC curvature and SHAP values (Fig. 5e), suggesting that increased LC curvature is linked to accelerated progression. The study of Ha et al. had a similar observation in POAG eyes where greater baseline LC curvature was associated with faster progression rate.[33] Similarly, Tan and colleagues showed that deeper and more curved NT curvature is linked with advanced POAG.[34] Furthermore, the study by Wanichwecharungruang et al. reported correlations between visual field MD values with LC thickness and depth, reinforcing the clinical relevance of LC morphology in tracking functional deterioration in PACG.[35] Our findings are consistent with these reports and further underscore the importance of ONH structural changes in predicting the progression of PACG.

We chose VFI slope as the primary measure of visual field progression, rather than the MD slope. While both VFI and MD are widely used in the literature to assess glaucoma progression, several studies have highlighted that progression rates vary significantly among patients, with the rate of change being one of the most reliable predictors of future deterioration.[36,37] Unlike MD, which can be confounded by media opacities, such as cataracts, VFI was specifically designed to mitigate these influences.[8,36] As cataract severity increases, MD may inaccurately suggest accelerated visual field loss, while post-cataract surgery, MD values often improve which complicates longitudinal assessments. In contrast, VFI is less sensitive to such changes and places greater emphasis on the central visual field, aligning more closely with ganglion cell damage, a key factor in glaucoma.[36] Moreover, the Humphrey Field Analyzer software incorporates the VFI slope to quantify the rate of progression, expressed as the percentage of visual field loss per year. Given its robustness and clinical relevance, we used VFI slope in this study to classify eyes as slow or fast progressors.

This study has several limitations that should be acknowledged. First, it was a hospital-based study with participants primarily of Chinese ethnicity, potentially limiting the generalizability of the findings to other populations. Second, patients with severe PACG were excluded from this analysis. While this may appear to limit generalizability, previous studies suggest that such patients contribute minimally to measurable VF progression. This is likely due to a "ceiling effect" where advanced baseline VF damage restricts the ability to detect further functional decline.[38] Third, the study used data from a single academic center and exclusively incorporated 24-2 Humphrey Visual Field tests, which may restrict the applicability of the machine learning model to broader clinical settings where different protocols are used. Fourth, only visual field and OCT imaging data were considered, while other relevant non-image factors such as age, gender, IOP, and treatment history were excluded; incorporating these variables could enhance model performance and interpretability in future research. Fifth, all OCT scans were acquired using a single imaging device. Given that image quality and segmentation accuracy can vary across machines, external validation using data from multiple devices is necessary to ensure robustness and generalizability of the findings. Finally, we utilized clinically well-established structural parameters of the ONH. In this context, deep learning-based methods could be employed to identify novel biomarkers, such as those proposed by Panda et al. and Sharma et al., which may demonstrate stronger correlations with disease progression.[39,40]

In summary, this study demonstrated that combining structural parameters from the ONH with functional parameters enhances the accuracy of progression prediction. The proposed methodology effectively identified key parameters associated with progression and revealed their correlations. This framework has the potential to support clinicians and researchers in gaining deeper insights into PACG progression and the contributing factors that should be considered.

**Acknowledgements**

We acknowledge funding from (1) the donors of the National Glaucoma Research, a program of the BrightFocus Foundation, for support of this research (G2021010S TableMJAG]); (2) National Medical Research Council, 20 Singapore (MOH-000372) (3) the NMRC-LCG grant 'TAckling & Reducing Glaucoma Blindness with Emerging Technologies (TARGET)', award ID: MOH-OFLCG21jun-0003 [AT/MJAG], (4) the Emory Eye Center (Emory University School of Medicine, Start-up funds, MJAG), (5) a Challenge Grant from Research to Prevent Blindness, Inc. to the Department of Ophthalmology at Emory

University, and (6) the NIH grant P30EY06360 to the Atlanta Vision Community (7) the National Eye Institute of the National Institutes of Health (NEI/NIH) - 1R01EY037299-01 (MJAG).

Supplementary material:

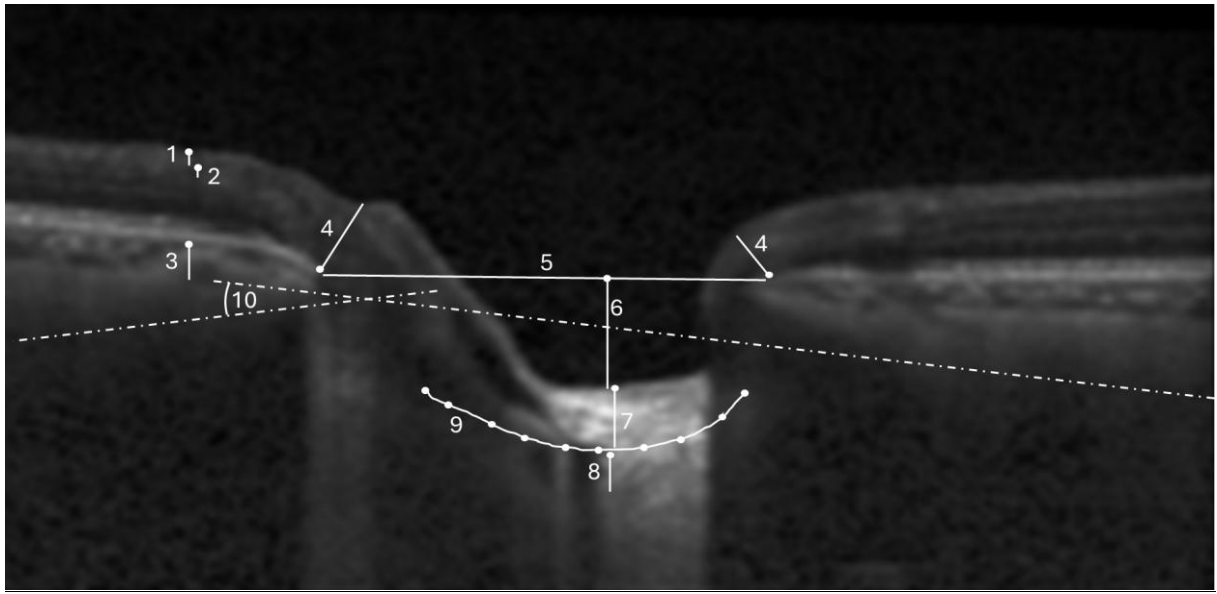

Figure A1: Structural parameters extracted from OCT images. 1: RNFL thickness, 2: GCL+IPL thickness, 3: Choroid thickness, 4: Minimum Rim Width, 5: Disc diameter, 6: Prelamina depth, 7: Prelamina thickness, 8: LC depth, 9: LC curvature, 10: Sclera angle

BMO_area: Bruch's Membrane Opening (BMO) area

lc_GSI: LC Global Shape Index

lc_curvature: Lamina Cribrosa curvature

lc_depth: LC depth

prelamina_depth: Prelamina depth

prelamina_thickness: Prelamina thickness

sclera_angle: Sclera angle

RNFLT_T: Average RNFL thickness at temporal region

RNFLT_ST: Average RNFL thickness at superior-temporal region

RNFLT_S: Average RNFL thickness at superior region

RNFLT_SN: Average RNFL thickness at superior-nasal region

RNFLT_N: Average RNFL thickness at nasal region

RNFLT_IN: Average RNFL thickness at inferior-nasal region

RNFLT_I: Average RNFL thickness at inferior region

RNFLT_IT: Average RNFL thickness at inferior-temporal region

GCCT_T: Average GCL+IPL thickness at temporal region

GCCT_ST: Average GCL+IPL thickness at superior-temporal region

GCCT_S: Average GCL+IPL thickness at superior region

GCCT_SN: Average GCL+IPL thickness at superior-nasal region

GCCT_N: Average GCL+IPL thickness at nasal region

GCCT_IN: Average GCL+IPL thickness at inferior-nasal region

GCCT_I: Average GCL+IPL thickness at inferior region

GCCT_IT: Average GCL+IPL thickness at inferior-temporal region

MRW_T: Minimum Rim Width at temporal region

MRW_ST: Minimum Rim Width at superior-temporal region

MRW_S: Minimum Rim Width at superior region

MRW_SN: Minimum Rim Width at superior-nasal region

MRW_N: Minimum Rim Width at nasal region

MRW_IN: Minimum Rim Width at inferior-nasal region

MRW_I: Minimum Rim Width at inferior region

MRW_IT: Minimum Rim Width at inferior-temporal region

Dependency plot of variables with SHAP values for slow and fast progression:

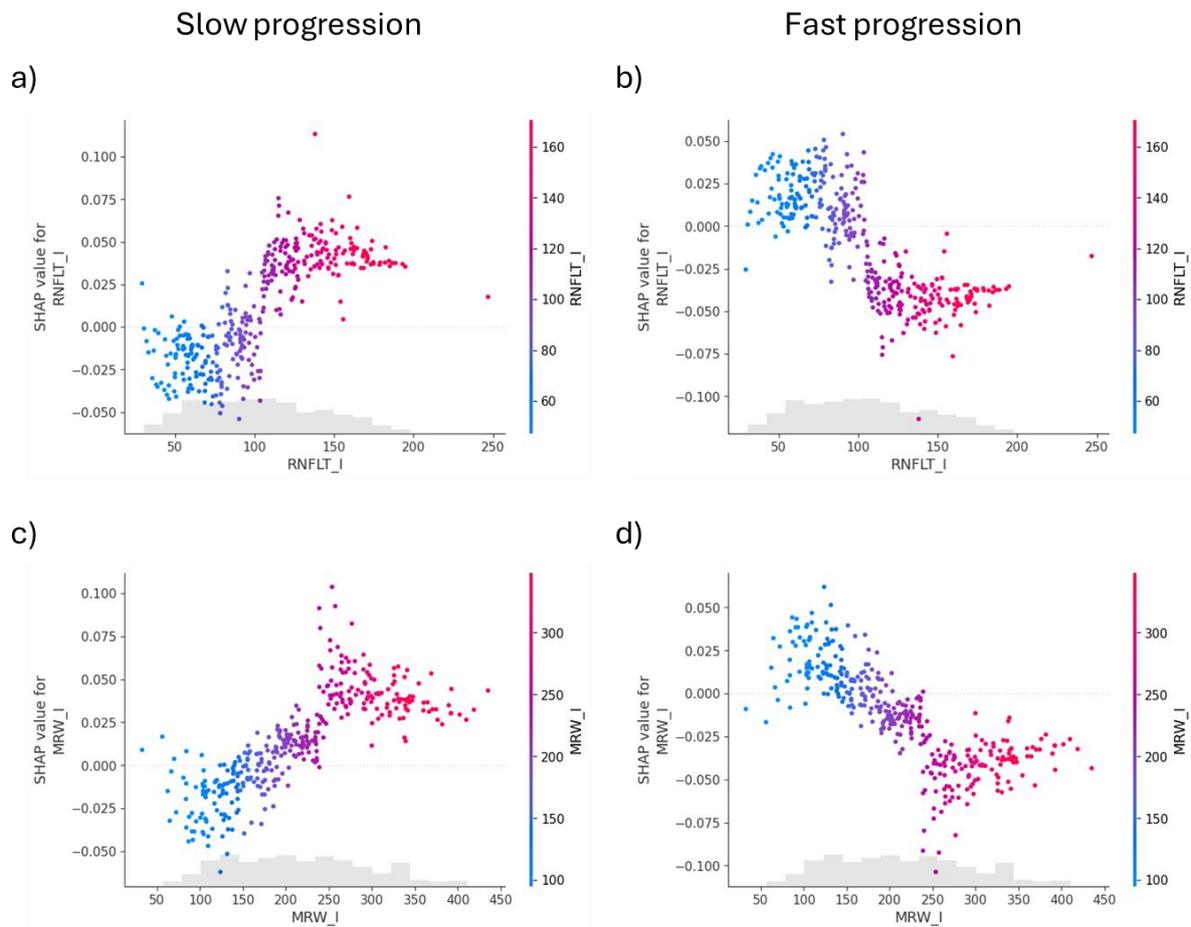

*Figure A2: a, b) Variation in SHAP values with average RNFL thickness at inferior region: a) variation during slow progression, b) variation during fast progression. c, d) Variation in SHAP values with MRW at inferior region: c) variation during slow progression, d) variation during fast progression.*